# Si-Substituted MAX Phases and *In-Situ* Formation of Si-coated MXene Composites via Chlorosilane Etching

*Xudong Wang[a, b, c]*, *Qian Fang[a, b, c]*, *Mian Li[a, c]*, *Zhifang Chai[a, c]*, *Qing Huang\*[a, c]*

[a] X. Wang, Q. Fang, Prof. M. Li, Prof. Z. Chai, Prof. Q. Huang, Zhejiang Key Laboratory of Data-Driven High-Safety Energy Materials and Applications, Ningbo Key Laboratory of Special Energy Materials and Chemistry, Ningbo Institute of Materials Technology and Engineering, Chinese Academy of Sciences, Ningbo, 315201, China.
E-mail: huangqing@nimte.ac.cn

[b] X. Wang, Q. Fang, University of Chinese Academy of Sciences, Beijing, 100049, China.

[c] X. Wang, Q. Fang, Prof. M. Li, Prof. Z. Chai, Prof. Q. Huang, Qianwan Institute of CNiTECH, Ningbo, 315336, China.

**Abstract**

Silicon-based MAX phases are a promising class of layered ceramics with superior thermal and chemical stability. However, their synthesis remains challenging due to inherent thermodynamic instability at high temperatures. Herein, we develop a general top-down strategy to synthesize a broad family of Si-substituted MAX phases (M = Ti, V, Nb, Ta, Cr; X = C, N) by reacting Al-based MAX precursors with $SiCl_4$ vapor. This approach not only circumvents traditional high-temperature limitations but also enables precise A-site defect engineering, resulting in phases with controlled vacancy concentrations (e.g., $Nb_2Si_{3/4}C$ and $Nb_2Si_{1/2}C$). Furthermore, we introduce a redox potential-based model that rationalizes the reaction pathway. Using $Ti_{n+1}AlX_n$ etched with $SiCl_4$ as an example, the process simultaneously forms Cl-terminated MXene ($M_{n+1}X_nCl_2$) and amorphous nano-Si, enabling the one-step synthesis of Si-coated MXene composites. This methodology provides new avenues for designing advanced MAX phases and MXene-based hybrids with tailored functionalities for applications in energy storage and catalysis.

**Introduction**

The family of layered ternary ceramics known as MAX phases and their two-dimensional derivatives, MXenes (obtained by selectively etching the A layer from MAX phases), have attracted significant attention due to their exceptional properties and broad application potential.[1-8] These layered materials are generally denoted as $M_{n+1}AX_n$ [1](MAX phases) and $M_{n+1}X_nT_x$[2] (MXenes), where M represents an early transition metal, A is an element from the main group, X denotes carbon or nitrogen, and $T_x$ stands for surface functional groups such as chalcogen or halogen elements.

To date, over 300 MAX phases have been successfully synthesized.[1] Among the diverse family of MAX phases, Si-based MAX phases (A = Si) have attracted particular interest due to their excellent thermal conductivity,[9] corrosion resistance,[10, 11] high-temperature plasticity and creep resistance.[12, 13] For example, compared to Al-MAX phases, $Ti_3SiC_2$ exhibits superior chemical inertness against lead-bismuth eutectic (LBE) alloys, positioning it as a promising candidate material for molten salt lead-bismuth reactors.[14, 15] Its remarkable resistance to acids and alkalis also suggests great potential as an electrochemical anode for chlorine generation in hydrochloric acid electrolytes.[16, 17](**patent**) Furthermore, the excellent interfacial compatibility between $Ti_3SiC_2$ and silicon carbide (SiC) makes it an ideal contact material for SiC-based electronic devices.[18, 19] (**patent**) Unfortunately, the vast potential of Si-based MAX phases remains largely untapped, as research to date has been almost exclusively confined to $Ti_3SiC_2$ and higher-order Ti-based variants ($Ti_4SiC_3$[20]). This severe limitation impedes systematic studies on their structure-property relationships, property optimization via composition regulation, and exploration of broader applications.

Conventionally, MAX phases are typically synthesized using bottom-up powder metallurgy methods at temperatures exceeding 1000 °C.[1] However, a fundamental challenge is that theoretical studies consistently indicate that Si-based MAX phases are inherently thermodynamically unstable due to the formation of competitive phases during high-temperature synthesis.[1, 21] This instability has greatly hindered their exploration. Although there have been successful reports of A-site solid solutions involving main group elements and Si (e.g., $Cr_2(Al_{1-x}Si_x)C$, $x \leq 0.03$,[22] $Zr_3(Al_{1-x}Si_x)C_2$,

$x < 0.1$[23] and $Ti_3(Si_{1-x}Ge_x)C_2$, $x = 0.25$[24]), the low solid solubility of Si is insufficient to significantly modulate intrinsic physicochemical properties. An alternative approach is the top-down method, which enables stable synthesis of MAX phases at lower temperatures, effectively avoiding the formation of high-temperature competitive phases and reducing the thermodynamic barriers. The effectiveness of this approach has been demonstrated by the substitution of late transition metals (Fe,[25] Co,[25] Ni,[25] Cu,[26] Zn[27]) and even noble metals (Au,[28] Ir,[28] Pt[29]) into the A-site of MAX phases within Lewis acid molten salts. Currently, synthesizing Si-based MAX phases with M-sites occupied by other transition metals remains a significant challenge.

Herein, we propose a top-down strategy to synthesize novel Si-substituted MAX phases ($M_{n+1}SiX_n$, where M = Ti, V, Nb, Ta, Cr and X = C or N) using $SiCl_4$ as a gaseous Lewis acid. The mechanism, elucidated through redox potential analysis, not only provides a rational pathway for synthesizing these previously elusive phases but also enables precise A-site defect engineering and MXene fabrication **(Figure 1)**. Notably, Si-coated MXene composites can be constructed *in-situ* via a one-step process. This approach may offer new perspectives for re-evaluating the applications of MAX phases and MXenes in catalysis and energy storage.

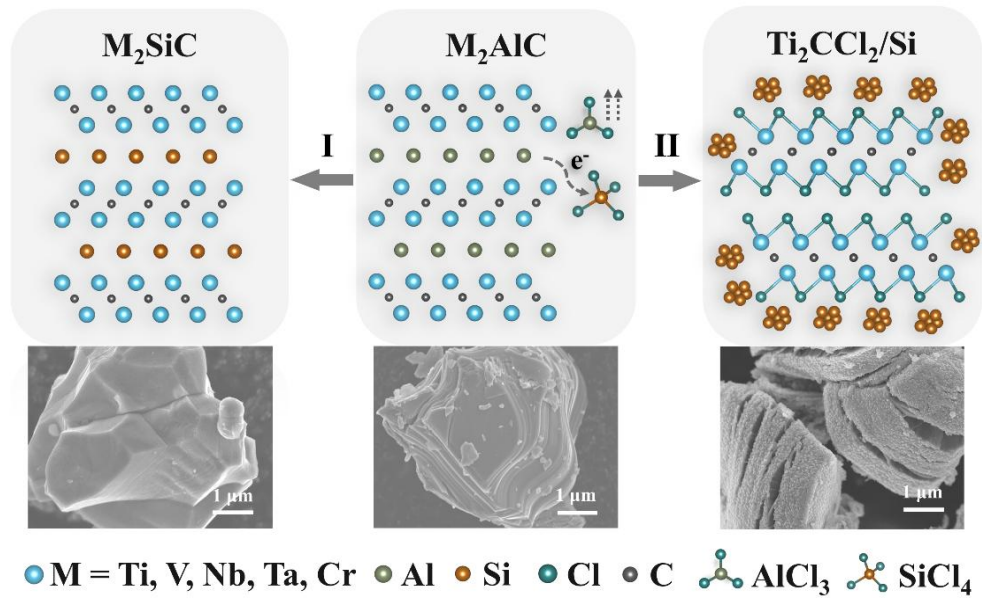

**Figure 1**. Schematic illustration of MAX phase etching by $SiCl_4$. Two distinct reaction pathways are presented: Path I leads to the formation of Si-substituted MAX phases,

whereas Path II yields a Si-coated MXene composites. The corresponding morphology image is presented below the atomic schematic diagram.

**Results and Discussion**

**Figure 1** illustrates the mechanism and reaction pathways. At the reaction temperature, the Al-MAX phase intentionally reacts with $SiCl_4$ vapor. In this process, the $Si^{4+}$ cation from $SiCl_4$ acts as a Lewis acid to seize electrons from the Al atoms in the MAX phase, yielding Si and $AlCl_3$. The reaction then proceeds via either Pathway I or II, yielding different products depending on the $SiCl_4$ quantity. A 4:3 molar ratio of Al-MAX to $SiCl_4$ directs the reaction along Pathway I, where the *in-situ* generated Si incorporates into the $M_6X$ interlayers, resulting in the novel Si-substituted MAX phases. Conversely, a 4:5 ratio promotes Pathway II, leading to the deposition of the generated Si onto the surface and thus the formation of Si-coated MXene composites.

**Si-substituted MAX Phases**

Conventionally, Si-based phases (e.g., $Ti_3SiC_2$) are synthesized through a bottom-up approach by sintering elemental powders.[30] However, theoretical calculations indicate that most Si-based MAX phases are unstable during conventional bottom-up sintering processes under thermodynamic equilibrium,[1] which severely limits their synthesis. To circumvent this limitation, we developed a top-down structural templating approach **(Figure 1)** to successfully fabricate several novel Si-substituted MAX phases. The Si-substituted MAX phases can be obtained by reacting the precursor MAX phase with $SiCl_4$ at a controlled molar ratio. The composition and structure of the as-synthesized products were systematically characterized to confirm the successful substitution of Si for Al. **Figure 2a** shows the X-ray diffraction (XRD) patterns of both the precursor $Ti_2AlC$ and the reaction product $Ti_2SiC$. Comparative XRD analysis reveals a distinct shift toward higher angles for $Ti_2SiC$ relative to $Ti_2AlC$, indicating a reduced lattice parameter. This phenomenon is particularly pronounced for the (002) diffraction planes **(the inset in Figure 2a)**, which can be attributed to the smaller atomic radius of Si compared to Al. The lattice parameters of $Ti_2SiC$ obtained by Rietveld refinement were determined to be $a$ = 0.3047 nm and $c$ = 1.2804 nm **(Figure 2b and Table S2)**, representing a significant contraction compared to those of the precursor ($a$ = 0.3058 nm, $c$ = 1.3649 nm).[31] The atomic coordinates of the constituent elements were refined as Ti (1/3, 2/3, 0.0938), Si (1/3, 2/3, 3/4) and C (0, 0, 0). The lattice parameters and $z$-coordinate of Ti ($z_M$) derived from XRD measurements show good agreement with theoretical calculations ($a$ = 0.3035 nm, $c$ = 1.2803 nm and $z_M$ = 0.0920).[32] Figure 2c presents a scanning electron microscopy (SEM) micrograph of Ti$_2$SiC, revealing a layered structure that is characteristic of MAX phases. The crystal structure was further confirmed by selected-area electron diffraction (SAED), which yielded lattice parameters (a = 0.3051 nm, c = 1.2801 nm, **Figure S1b**) consistent with XRD refinement. Atomic-resolution scanning transmission electron microscopy (STEM) micrographs unambiguously reveal the hallmark mirrored zig-zag atomic configuration of MAX phases (**Figure 2d**). Atomic number contrast analysis identifies brighter spots as Ti atoms, with darker spots between Ti$_6$C octahedra corresponding to Si atoms. This elemental distribution is further validated by high-resolution energy-dispersive X-ray spectroscopy (EDS) mapping, which confirms the homogeneous presence of Ti, Si, and C without detectable Al **(Figure 2e, Figure S1c-d)**. In conclusion, this top-down approach circumvents the thermodynamic instability issues associated with the direct sintering of elemental powders, thereby providing a viable new pathway for the synthesis of a series of novel Si-substituted MAX phases.

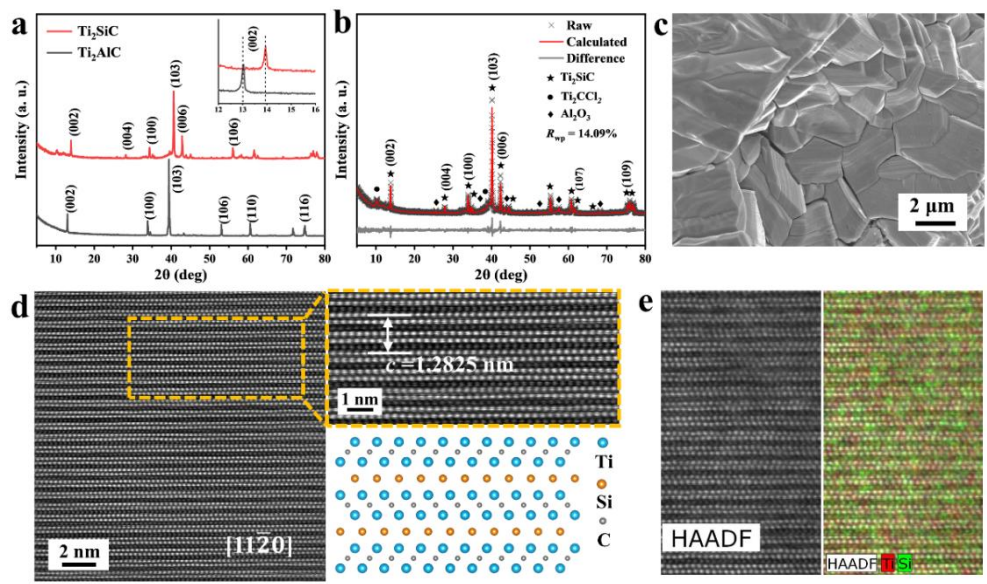

**Figure 2**. Structural characterization of Ti$_2$SiC. (a) XRD patterns comparing Ti$_2$SiC and its parent phase Ti$_2$AlC; the inset highlights the (002) diffraction peak. (b) Rietveld refinement of the XRD pattern for Ti$_2$SiC. (c) SEM micrograph of Ti$_2$SiC showing its typical layered morphology. (d) Atomic-resolution STEM image viewed along the [11$\bar{2}$0] zone axis with a corresponding atomic structure schematic. (e) Elemental distribution revealed by atomic-resolution EDS mapping, indicating Ti and Si signals.

To demonstrate the generality of our top-down synthesis approach, we extended it to fabricate Si-substituted MAX phases with different M-site elements, including Ti$_2$SiN, V$_2$SiC, Nb$_2$SiC, Ta$_2$SiC, Cr$_2$SiC, and Ta$_4$SiC$_3$ (**Table S1**). **Figure 3a-f** presents the XRD patterns and corresponding Rietveld refinement results of these phases, confirming that the predominant products are phase-pure Si-substituted MAX phases with only minor carbide or nitride. The lattice parameters and respective atomic coordinates of component elements in those Si-substituted MAX phases were successfully determined via XRD Rietveld refinement method, showing good agreement with theoretical calculations (**Table S2**). As expected from the smaller atomic radius of Si, all the resultant Si-substituted MAX phases exhibit reduced lattice constants compared to their parent Al-MAX phases, a conclusion further corroborated by overlaid XRD patterns (**Figure S2-S4**). Further characterization by scanning electron microscopy (SEM) revealed that all products exhibit the typical layered morphology of MAX phases (**Figures S2-S4**). EDS analysis confirmed the near-complete replacement of Al by Si, with the M/Si atomic ratios matching those expected for the M$_{n+1}$SiC$_n$ structures. It is noteworthy that accurately determining the Ta/Si ratio was challenging due to the overlapping Ta M-line (~1.71 keV) and Si Kα peak (~1.74 keV). Despite this analytical challenge, the concomitant disappearance of Al and the consistent shift of the (002) peak across all samples provide conclusive evidence for the successful synthesis of Ta$_{n+1}$SiC$_n$.

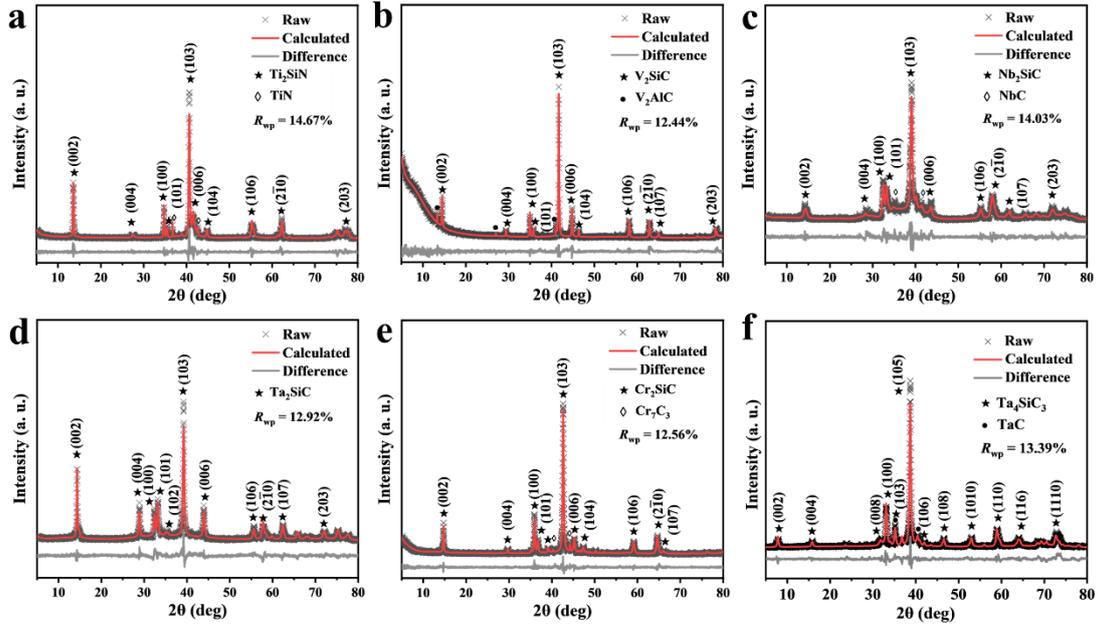

**Figure 3.** Rietveld refinement XRD patterns of Si-substituted MAX phases: (a) Ti$_2$SiN, (b) V$_2$SiC, (c) Nb$_2$SiC, (d) Ta$_2$SiC, (e) Cr$_2$SiC, and (f) Ta$_4$SiC$_3$.

**Defect Engineering of Si-substituted MAX phases**

$$2\ M_{n+1}AlC_n + 3\ BCl_2 = M_{n+1}BC_n + AlCl_3{\uparrow} + B,\ (B = Zn, Cu, Co, Fe) \quad (1)$$

$$4\ M_{n+1}AlC_n + 3\ SiCl_4(g) = 4\ M_{n+1}Si_{3/4}C_n + 4\ AlCl_3{\uparrow} \quad (2)$$

$$2\ Nb_2ZnC + SiCl_4(g) = 2\ Nb_2Si_{1/2}C + 2\ ZnCl_2 \quad (3)$$

Lewis acid cations typically employed for isomorphous substitution in MAX phases are divalent (e.g., Zn$^{2+}$ from ZnCl$_2$, Fe$^{2+}$ from FeCl$_2$), as exemplified by Equation (1). Governed by charge and mass conservation, these chemical reactions generally yield derivative MAX phases with saturated A-sites (e.g., Ti$_3$ZnC$_2$[27], Ta$_2$FeC[25]), often accompanied by detectable residual B metal (e.g., Zn, Fe) in the final products. In stark contrast, the tetravalent nature of Si$^{4+}$ (from SiCl$_4$) dictates a different outcome. Theoretical considerations based on Equation (2) predict that the resulting Si-substituted MAX phases should inherently exhibit A site vacancy ($V_A$), characterized by a vacancy fraction of 1/4. To experimentally verify this prediction, Nb$_2$SiC was synthesized as an example and analyzed. Rietveld refinement of the XRD patterns revealed a Si occupancy of merely 0.75 at the A-site. This finding was also corroborated by EDS analysis, which yielded a Nb/Si atomic ratio of 2.43, corresponding to a Si occupancy of ~0.73. The excellent agreement between these techniques unambiguously confirms the presence of $V_A$ at a concentration of ~25 %, in full consistency with the theoretical prediction. This phenomenon was further generalized across other Si-substituted MAX phases **(Table S3)**, confirming its fundamental nature.

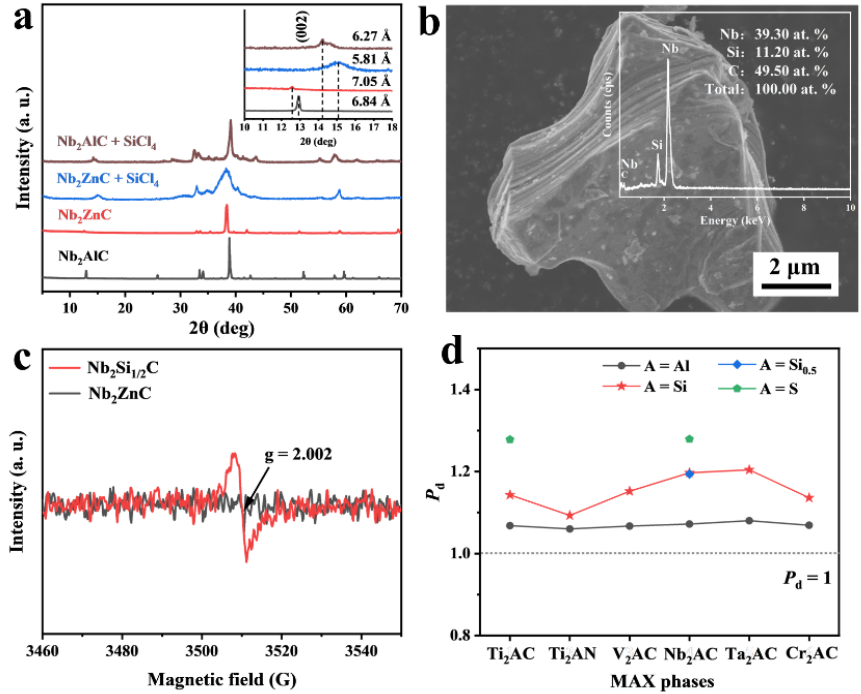

**Figure 4.** (a) XRD patterns of $Nb_2Si_{1/2}C$ compared with $Nb_2AlC$, $Nb_2ZnC$, and $Nb_2SiC$. (b) SEM image showing the morphology of $Nb_2Si_{1/2}C$ with corresponding EDS point analysis. (c) Electron spin resonance (ESR) spectra of $Nb_2Si_{1/2}C$ and $Nb_2ZnC$. (d) $P_d$ values of MAX phases with different A-site elements.

Encouraged by this result, we aimed to push the boundaries further by targeting a phase with an even higher $V_A$ concentration in MAX phases. By reacting precursor $Nb_2ZnC$ with $SiCl_4$ **[Equation (3)]**, we predicted the formation of a $Nb_2SiC$ phase with a remarkable 1/2 vacancy fraction at the A-site **(Figure S5)**. The (002) peak in the XRD pattern of the product (**Figure 4a**) shifted to a higher angle compared to the phase with ~25 % A-site vacancies [**from Equation (2)**]. This shift indicates a contraction of the interlayer spacing, consistent with a higher density of vacancies. Crucially, despite this extreme $V_A$ concentration, the structure retained its characteristic layered architecture, as vividly demonstrated by the SEM image in **Figure 4b**. Quantitative analysis determined the A-site occupancy to be 0.475 (XRD) and 0.57 (EDS), both values confirming the achievement of an exceptionally high $V_A$ concentration close to 50 %. To further solidify these findings, we employed electron paramagnetic resonance (EPR) spectroscopy. An intense EPR signal was observed at a g-value of 2.002 (**Figure 4c**), which can be directly attributed to Si vacancy centers. This provides definitive spectroscopic evidence for the successful creation of a MAX phase with approximately half of its A-sites vacant. The influence of $V_A$ on the distortion of $M_6A$ trigonal prisms ($P_d$) in MAX phases was evaluated using the lattice parameters and the z-coordinate of the M ($z_M$)[33] (**Table S4**):

$$P_d = \frac{1}{\sqrt{\frac{1}{3} + (\frac{1}{4} - z_M)^2 (\frac{c}{a})^2}}$$

The $P_d$ has an ideal value of 1.0 for close-packed hard spheres in 211 MAX phase structure. A value exceeding 1 indicates compression along the *c*-axis.[34] As shown in **Figure 4d**, the $P_d$ values for Al-MAX phases are close to 1.0, whereas those for S-MAX phases are consistently higher (~1.3). This trend is similar to that reported for $M_2SC$ (M = Zr, Hf)[35] and originates from their strongly covalent M–A bonds. In contrast, the M–A bonding in Al/Si-based MAX phases is typically metallic and weaker. Notably, the Si-substituted MAX phases in this work display $P_d > 1.1$, even exceeding 1.2, which may be attributed to $V_A$-induced interlayer distortion causing *c*-axis contraction.

The successful synthesis of such a $V_A$-rich phase represents a significant advance in the structural engineering of MAX phases. While double-layer A-site configurations (e.g., $Mo_2Ga_2C$,[36] $Ti_3Au_2C_2$[28] and $Nb_2Bi_2C$[29]) are known, the deliberate creation of these unsaturated, $V_A$-rich MAX phases is rare. We propose that these $V_A$ in MAX phases, introduced via defect engineering, could serve as highly reactive centers, opening promising avenues for applications in catalysis and energy storage.

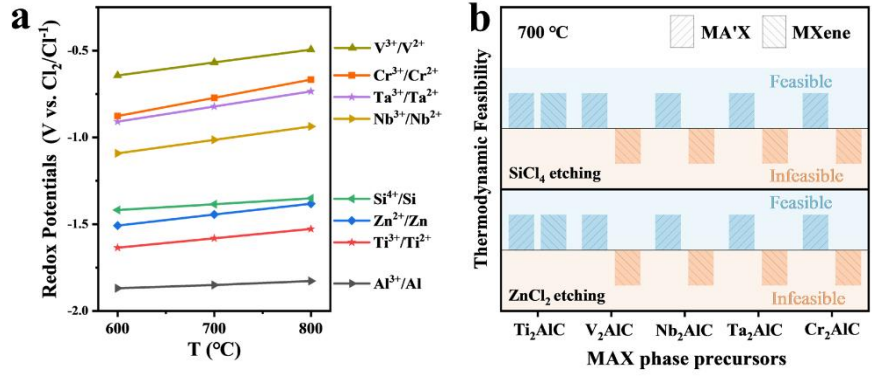

**Figure 5.** (a) Redox potentials (V vs. $Cl_2/Cl^-$) of the $Si^{4+}/Si$, $M^{3+}_{(MXene)}/M^{2+}_{(MAX)}$ and $A^{n+}_{(ACl_n)}/A_{(MAX)}$ redox couples as a function of temperature. (b) Assessment of thermodynamic feasibility for reactions between gaseous $SiCl_4$ and molten $ZnCl_4$ with various MAX phases at 700 °C, based on their redox potentials, for predicting reaction products.

**Formation Mechanism and *In-Situ* Si-coated MXene Composites**

$$8\ Ti_2AlC + 6\ SiCl_4(g) = 8\ Ti_2C + 8\ AlCl_3\uparrow + 6\ Si \quad (4)$$

$$8\ Ti_2C + 5\ Si + SiCl_4(g) = 2\ Ti_2CCl_2 + 6\ Ti_2SiC \quad (5)$$

$$8\ Ti_2AlC + 7\ SiCl_4(g) = 6\ Ti_2SiC + 2\ Ti_2CCl_2 + 8\ AlCl_3\uparrow + Si \quad (6)$$

$$4\ Ti_2AlC + 5\ SiCl_4(g) = 4\ Ti_2CCl_2 + 4\ AlCl_3\uparrow + 5\ Si \quad (7)$$

Combined XRD refinement, EDS analysis, and STEM imaging revealed the absence of Si vacancies in the product obtained from the reaction of $Ti_2AlC$ with $SiCl_4$ (**Figures 2d, S1c and Table S3**), contradicting the initial hypothesis derived from Equation (2). Moreover, even when stoichiometric amounts of reactants were used according to Equation (2), the resulting Si-substituted MAX product was consistently accompanied by $Ti_2CCl_2$ as a byproduct **(Figure 2b)**. This behavior starkly contrasts with that of non-Ti-based MAX parent phases (e.g., $Nb_2AlC$), which produced phase-pure Si-substituted MAX phases products without detectable MXene, even under supersaturated reaction conditions (elevated $SiCl_4$ concentrations or temperatures).

To unravel this anomaly, we turned to a thermodynamic framework based on redox potentials. The transformation between MAX phases typically involves only the oxidation of the A-site element and the insertion of a new A-element. In contrast,

MXene formation requires additional oxidation of the M-site elements to higher oxidation states.[27, 29] We therefore calculated the redox potentials (V vs. $Cl_2/Cl^-$) for relevant couples. Beyond the established couples for the A-site ($A^{n+}/A$) and silicon ($Si^{4+}/Si$) used in MAX phase etching predictions,[37] we innovatively incorporated the previously unreported M-site couples ($M^{3+}/M^{2+}$) within MAX phases (**Table S5 and Figure 5a; see Supporting Information for details**). Our calculations over the temperature range of 600-800 °C revealed a decisive redox potential order: $E(V^{3+}/V^{2+}) > E(Cr^{3+}/Cr^{2+}) > E(Ta^{3+}/Ta^{2+}) > E(Nb^{3+}/Nb^{2+}) > E(Si^{4+}/Si) > E(Zn^{2+}/Zn) > E(Ti^{3+}/Ti^{2+}) > E(Al^{3+}/Al)$ (**Figure 5a**). Critically, the $E(Si^{4+}/Si)$ is higher than $E(Ti^{3+}/Ti^{2+})$ but lower than those of Nb, Ta, Cr, and V. This hierarchy implies that $SiCl_4$ is thermodynamically capable of oxidizing both the A-site (Al) and the M-site (Ti) in $Ti_2AlC$, creating a thermodynamic competition between Si-substituted MAX phases formation and MXene etching. For other M elements (e.g., Nb), $SiCl_4$ can oxidize only the A-site, leading exclusively to a clean substitution reaction. Furthermore, the formation of a $V_A$-rich $Ti_2SiC$ is energetically unfavorable due to the penalties associated with vacancy formation and lattice distortion. To minimize the total energy, the system avoids this pathway. Instead, a portion of the M-site (Ti) is selectively oxidized, leading to the stable MXene $Ti_2CCl_2$. Thus, this process effectively prevents the consumption of Si atoms, thereby enabling all available Si to occupy the A-sites and form the $V_A$-free $Ti_2SiC$. This mechanism is comprehensively described by Equations (4) to (6), where (4) and (5) represent the individual reactions, and (6) corresponds to the overall reaction. The predictive power of this redox potential framework is further underscored by the order $E(Zn^{2+}/Zn) > E(Ti^{3+}/Ti^{2+}) > E(Al^{3+}/Al)$, which correctly predicts that $ZnCl_2$ can either form $Ti_2ZnC$ or $Ti_2CCl_2$ from etching $Ti_2AlC$, a phenomenon validated in prior work[27].

Our innovative integration of M-site redox potentials into the analysis of MAX phase transformations provides a novel and powerful strategy for predicting reaction pathways. As illustrated in **Figure 5b**, $Ti_2AlC$ can be transformed into both a substituted MAX phase and MXene when reacting with $SiCl_4$, whereas other MAX phases (where M = V, Nb, Ta, Cr) only convert into substituted MAX phases. A similar trend is observed in reactions between M$_2$AlC and ZnCl$_2$, as previously reported and explained.[27] This conceptual advance not only explains our present results but also paves the way for a rational design strategy. We envision that a comprehensive database of redox potentials for M, A and X elements could be established in the future, which would be instrumental in selectively etching challenging MAX phases and guiding the synthesis of novel structures.

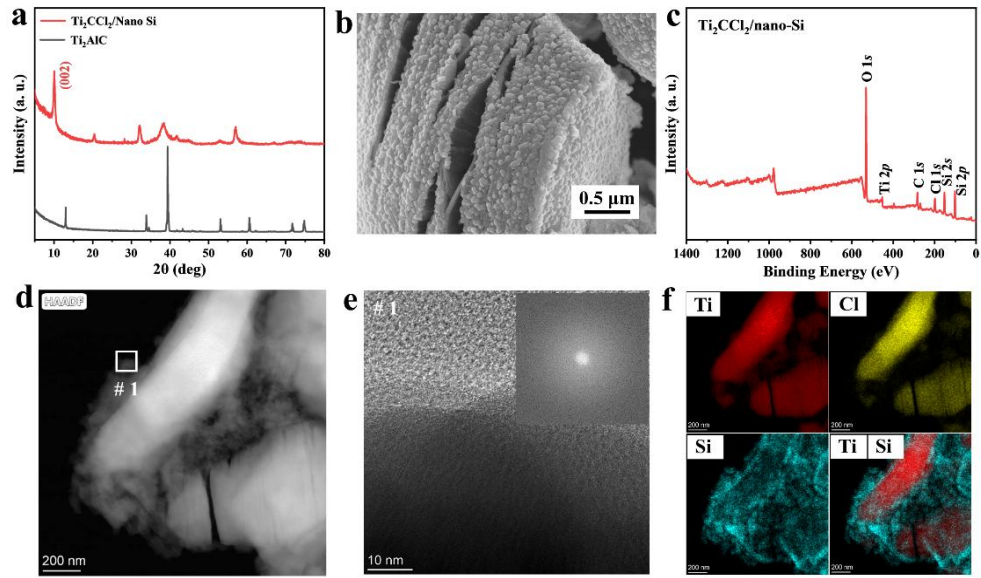

**Figure 6**. Characterization of Si-coated Ti$_2$CCl$_2$ composite. (a) XRD patterns of Si-coated Ti$_2$CCl$_2$ and Ti$_2$AlC, confirming the formation of Cl-terminated MXene. (b) Morphology of Si-coated Ti$_2$CCl$_2$ showing uniform distribution of dense Si nanoparticles on the MXene surface. (c) XPS analysis of Si-coated Ti$_2$CCl$_2$ composite. (d-f) TEM analysis: HAADF image (d), HRTEM image showing the amorphous/disordered state of nano-Si (e) and corresponding elemental mapping revealing encapsulation of MXene by Si particles (f).

Guided by the predictive power of our redox potential framework, we selectively etched Ti$_2$AlC using an excess SiCl$_4$ vapor to achieve the synthesis of MXene [Equation (7)]. This etching behavior fundamentally differs from conventional methods employing ZnCl$_2$ or CuCl$_2$, which yield crystalline metallic byproducts (e.g., micron-sized Zn or Cu) that are easily detectable by XRD.[27, 37] In contrast, our process yields primarily amorphous silicon, as confirmed by the absence of crystalline Si peaks in the XRD pattern (**Figure 6a**). The pattern instead is dominated by the characteristic (002) peak of $Ti_2CCl_2$ MXene at 10.2°, providing unequivocal evidence of its successful synthesis. X-ray photoelectron spectrum (XPS) analysis further corroborates the formation of $Ti_2CCl_2$ and the presence of silicon (**Figure 6c and Figure S6**). The Ti 2p spectrum exhibits peaks at 454.73 eV and 456.19 eV, corresponding to the Ti–C ($2p_{3/2}$) environments,[27, 38] and a peak at 457.8 eV is assigned to the Ti–Cl ($2p_{3/2}$).[39] The positive chemical shift of the Ti–Cl peak relative to the Ti–C peaks is consistent with the higher oxidation state of Ti bonded to Cl. The Cl 2p spectrum shows a distinct peak at 199.82 eV, further confirming the presence of Ti–Cl ($2p_{3/2}$)[39]. No signal is detected in the Al 2p spectrum, indicating complete etching of Al. Additionally, the Si 2p spectrum reveals peaks at 101.63 eV and 103.2 eV, attributable to Si–O and $SiO_2$.[40] This superficial oxidation likely occurred on the freshly exposed nano-Si during the $AlCl_3$ removal process.

Morphological and elemental analysis using SEM-EDS revealed a dense distribution of silicon nanoparticles (approximately 100 nm in size) adsorbed on the MXene surface (**Figure 6b and Figure S8**), with EDS result confirming the presence of Si (**Figure S7 and Figure S8c-d**). TEM analysis showed that the composite structure consists of nanoscale silicon enveloping the MXene matrix (Si-coated $Ti_2CCl_2$) (**Figure 6d–f**). High-resolution transmission electron microscopy (HRTEM) provides direct evidence of Si layer coating the $Ti_2CCl_2$ MXene (**Figure S9b**). Images from Si-rich regions exhibit a typical mottled contrast (**Figure 6e**), and the corresponding Fast Fourier Transform (FFT) patterns show only diffuse rings (**the inset in Figure 6e**), collectively confirming the amorphous nature of the silicon in the product. The versatility of our redox-guided approach is further demonstrated by its successful application to $Ti_3AlC_2$, from which the anticipated Si-coated $Ti_3C_2Cl_2$ composite was similarly obtained (**Figure S10**).

While Si-coated MXene composites are desirable for applications like lithium-ion battery anodes,[41, 42] their synthesis typically relies on multi-step mixing procedures. Notably, our method enables the one-step, *in-situ* synthesis of such composites via a single vapor-phase reaction. This approach presents a significant simplification and a potential route for scale-up, which could be beneficial for future energy storage applications.

**Conclusion**

In conclusion, we have developed a general top-down strategy using $SiCl_4$ vapor to synthesize a series of novel Si-substituted MAX phases ($M_{n+1}SiX_n$, M = Ti, V, Nb, Ta, Cr; X = C, N), overcoming the long-standing thermodynamic instability that limited their conventional synthesis. By introducing a novel redox potential model that accounts for both A-site and M-site elements, we established a predictive framework for reaction pathways, rationalizing the selective formation of phase-pure Si-substituted MAX phases or Si-coated MXene composites. A key achievement is the controlled creation of $V_A$ (~25 %), inherent to this synthesis mechanism, with the possibility to extend the $V_A$ concentration to nearly 50 % using pre-substituted MAX precursors. These $V_A$ induce significant structural distortion, as confirmed by XRD refinement and ESR spectroscopy, which is expected to strongly influence material properties. Furthermore, this approach enables the one-step *in-situ* synthesis of Si-coated MXene composites, a highly desirable architecture for energy storage applications. This work not only expands the family of MAX phases and enables precise defect engineering, but also provides a universal thermodynamic guideline for the rational design and synthesis of advanced layered materials and their derivatives.

**Supporting Information**

This document provides experimental details, materials specifications, synthesis procedures for Si-substituted MAX phases and *in-situ* Si-coated MXene composites, comprehensive XRD Rietveld refinement data, SEM/TEM/EDS analyses, XPS and EPR characterizations, and thermodynamic calculations of redox potentials for the involved reactions.

**Acknowledgements**

This study was supported financially by the National Natural Science Foundation of China (grant numbers U23A2093 and 12375279), the Ningbo Top-talent Team Program, and the Ten-Thousand Talents Plan of Zhejiang Province (grant number 2022R51007).